\documentclass[useAMS,usenatbib]{mn2e}
\usepackage{graphicx}

\title[Chaotic dynamics in DF Cyg]{Chaotic dynamics in the pulsation of DF~Cygni, as observed by \textit{Kepler}}
\author[E. Plachy et al.]{E. Plachy$^{1,2}$\thanks{E-mail:
eplachy@konkoly.hu}, A. B\'odi$^{1,2,3}$ and Z. Koll\'ath$^{4}$ \\\\
$^{1}$Konkoly Observatory, MTA CSFK, Konkoly Thege Mikl\'os \'ut 15-17, H-1121 Budapest, Hungary\\
$^{2}$MTA CSFK Lend\"ulet Near-Field Cosmology Research Group\\
$^{3}$Department of Experimental Physics and Astronomical Observatory, University of Szeged, Hungary\\
$^{4}$E\"otv\"os Lor\'and University, Savaria Department of Physics, Szombathely, Hungary\\
}

\begin{document}

\date{Accepted. Received; in original form }

\pagerange{\pageref{firstpage}--\pageref{lastpage}} \pubyear{2002}

\maketitle

\label{firstpage}

\begin{abstract}

Pulsations of RV Tauri-type variable stars can be governed by chaotic dynamics. However, observational evidence for this happening is usually hard to come by. Here we use the continuous, 4-year-long observations of the \textit{Kepler} space telescope to search for the signs of chaos in the RVb-type pulsating supergiant, DF Cygni. We use the Global Flow Reconstruction method to estimate the quantitative properties of the dynamics driving the pulsations of the star. The secondary, long-term light variation, i.e., the RVb phenomenon was removed in the analysis with the Empirical Mode Decomposition method. Our analysis revealed that the pulsation of DF Cyg could be described as a chaotic signal with a Lyapunov dimension of  $\sim$2.8. DF Cyg is only the third RV Tau star, and the first of the RVb subtype, where the nonlinear analysis indicates that low-dimensional chaos may explain the peculiarities of the pulsation.

\end{abstract}

\begin{keywords}
stars: variables: Cepheids -- stars: individual: DF Cyg -- chaos
\end{keywords}

\section{Introduction}
RV Tau stars are post-AGB supergiant stars representing the most luminous group of radial pulsators. They constitute the long period subclass of Type II Cepheids (P$>$20 days).

Two subtypes of RV Tau variables are distinguished: RVb stars show long-period large-amplitude variability in the mean brightness on top of the pulsation, while RVa stars do not show this phenomenon. Apart from the long-period variations, RVb light curves are similar to that of RVa stars. The origin of the secondary light variation is not fully understood yet. \citet{fokin} showed that intrinsic stellar processes such as pulsation or any thermal instabilities can not be responsible for the secondary variation. More recent studies explain RVb phenomenon with binarity: the mean brightness changes due to periodic obscurations by a circumbinary disc around the variable and its companion star \citep{vanwinckel,maas,pollard}. The possibility of interaction between the components has also been raised \citep{pollard2}.

The obscuration scenario seems to work for DF~Cyg as well. DF Cyg (KIC 7466053) is the only known RV~Tau star continuously observed from space in the \textit{Kepler} primary mission \citep{bodi}. Using \textit{Kepler} data, \citet{vega} showed that, when measured in fluxes, both the pulsation amplitudes and the mean brightness decrease by $\sim$90 percent during the long-period minimum, strongly supporting the disc occultation scenario. Adopting the idea of estimating the amplitude changes in flux units, recent investigations highlighted that many RVb stars show similar correlations between the pulsation amplitude variations and the long-term variability \citep{kissbodi}.  

RV Tau stars show alternations of deep and shallow minima in the light curve, that are thought to be the sign of the nonlinear phenomenon called period doubling. However, the alternation is not always recognizable due to the strong irregularity of the light variation that may signal more complex behaviour. Incidentally, many RV Tau stars also exhibit occasional interchanges in the order of low- and high-amplitude cycles \citep{ogle3,plachy2014a}, these features are observable in DF Cyg too. 

RV Tau stars are difficult to model, but theoretical calculations of shorter-period subtypes of Type-II Cepheids (W~Vir- and BL~Her-type stars) showed that both period doubling and chaos can be expected in their pulsation \citep{kovacs, moskalik1990, smolec2014, smolec2016}. Recent observations confirmed that period doubling is indeed a common feature in the period range of 16$<$P$<$20 days \citep[][Smolec et al., in prep.]{ogle4, plachy2017}. In contrast, detecting chaotic behavior and determining its fractal dimension in observational data is a real challenge due to the requirements on data length and quality. Only two RV~Tau stars have been analyzed with a nonlinear approach: R~Sct \citep{buchler1996} and AC Her \citep{kollath1998}. Both of them show low-dimensional chaos with Lyapunov dimensions of $\sim$3.1 in the case of R~Sct and 2.05-2.45 for AC~Her. 

In this paper we report on the dynamical analysis of the most continuous and accurate photometric data of an RV~Tau star available to date: the \textit{Kepler} light curve of DF~Cyg. DF~Cyg is classified as an RVb star, its pulsation period is $\sim$24.9 days while the length of the secondary, long-period variation is $\sim$780 days. This star is a popular target for amateur astronomers, and have been followed since the 1970s, but unfortunately the data collected at the AAVSO (American Association of Variable Star Observers) is too sparsely sampled to be useful in our analysis.   

Below we present two methods that have been applied in this study: the Global Flow Reconstruction (GFR) method to search and quantify the chaotic nature of the pulsation, and the Empirical Mode Decomposition (EMD) method to separate the long-period variation from the pulsation. Both methods are introduced in Section~\ref{met}. Details of the data preparation are given in Section~\ref{dat}. Analysis and results are discussed in Section~\ref{results}. We also provide summary and conclusions in Section~\ref{sum}.

\section{Methods}
\label{met}
Chaos may emerge in nonlinear dynamical systems. This seemingly random behavior has nothing to do with stochastic processes, but it is deterministic and has its own characteristic pattern. An important property of chaotic systems is the sensitivity to initial conditions: an infinitesimally small change at the starting point results in an entirely different future path. This property is measurable, and gives quantitative information about the system. The evolution of a dynamical system can be visualized in a phase space, the collection of possible system states, where evolving states trace a path drawing a trajectory. The dimension of the phase space is determined by the degrees of freedom of the system. If the motion reaches an equilibrium, the trajectory converges into a fixed point. A periodic motion represents a closed trajectory, called limit cycle. In contrast, a chaotic trajectory of a dissipative dynamical system is always an aperiodic, bounded  curve that tends to evolve towards a strange attractor. Strange attractors are not normal geometric objects, they have no integer, but fractal dimensions, and they typically occupy a small region in the phase space.

According to Takens' theorem \citep{takens} a single measured quantity is sufficient to reconstruct the attractor. In our case the only variable arising from the system is the brightness of the star. The dimension of the reconstructed phase space is called the embedding dimension, which must be larger than the fractal dimension of the system to avoid crossing of the trajectories. The reconstruction preserves the the essential mathematical properties of the original system like topology and Lyapunov exponents. The latter quantities describe the rates of exponential growth/contraction in the system by characterizing the rate of separation of infinitesimally close trajectories that can be different for different orientations in the phase space. A positive maximal Lyapunov exponent is considered as a definition of deterministic chaos. Lyapunov exponents are also used for the calculation of the fractal Lyapunov dimension. For general review of chaos theory and its applications in astrophysics we refer to \citet{Regev}. In the following sections we introduce the two fundamental tools applied in our analysis.

\subsection{The Global Flow Reconstruction method}
\label{gfr}
A powerful technique have been developed by \citet{Serre} for the special purpose of phase space reconstruction of stellar pulsation,
the Global Flow Reconstruction (hereafter GFR) method. The term "global flow"  denotes the dynamics of the hypothetical underlying system. The same
nonlinear analyzer tool was applied in the studies of R~Sct and AC~Her, and has been proven to be useful in searching for chaos in semiregular variables \citep{semireg}, W~Virginis models \citep{WVmod}, as well as in RR Lyrae models and observations \citep{plachy2013,plachy2014b}. Given the successful applications we adopted GFR in the analysis of DF Cyg.

The first step in the method is the transformation of the light curve into a data sequence with equal time spacing ${s(t_n)}$ and the production of the "delay vectors" $\mathbf{X}(t_n)={(s(t_n), s(t_n-\Delta),s(t_n-2\Delta),...,s(t_n-(d_e-1)\Delta))}$. $\Delta$ denotes the “time delay”, while $d_e$ is the embedding dimension of the reconstruction space. The method is optimized to detect low-dimensional chaos, so $d_e$ can be set to 4, 5 or 6. No higher embedding dimensions are implemented, because the number of variables becomes unmanageably high. 
We consider the dynamical system as an iterated map, the time evolution rules are given as algebraic equations, and the values are expressed as a function of the value of the previous step.
So we assume that there exists a map $\mathbf{F}$ that evolves the trajectory in time by connecting the consecutive points: $\mathbf{X}^{n+1}=\mathbf{F}(\mathbf{X}^n)$. We search $\mathbf{F}$ in polynomial form: $\mathbf{F(X)=}\sum_{k}C_{k}P_{k}\mathbf{(X)}$, where $P_{k}\mathbf{(X)}$ represent the polinomials of order up to $p$. The value of $p$ was fixed to 4. After we found $\mathbf{F}$, arbitrary long data sets can be produced that we call "synthetic signals". The length of the data plays an important role in the determination of the fractal dimension of the system. Our code computes the Lyapunov exponents $\lambda_i$, that give the divergence rates of infinitesimally close trajectories in each dimension ($i=1,2,3...$) of the embedding space: $|\delta \mathbf{Z}(t)| \approx e^{\lambda t} |\delta \mathbf{Z}_0|$, where $\mathbf{Z}_0$  is the initial separation. At least one Lyapunov exponent must be positive for a chaotic signal. Hundreds of cycles are required to calculate these values with sufficient accuracy, which are usually not available from observational data. Therefore we adopt the quantitative properties of the synthetic signals that show strong resemblance to the light curve instead. The calculation of the Lyapunov dimensions $d_L=K+\frac{1}{|\lambda_{K+1}|}\sum_{i=1}^{K}\lambda_i$ is also implemented in the method. 

The success of GFR rests upon the possibility of applying slight variations to the phase space trajectories. This can be achieved by adding a small amount of noise to the data which will be then subsequently smoothed. By varying the noise and smoothing parameters we construct dynamically very similar data sets for reconstruction. We usually use a large parameter space of noise, smoothing, and time delay values. Synthetic signals are created for each parameter set from the corresponding maps. Added noise is defined relative to the root mean square of the signal through the noise intensity parameter ($\xi$), which we usually set to be in the range of the observational noise of the data. The smoothing parameter ($\sigma$) denotes the maximum standard deviation from the fit in the cubic spline algorithm. These two parameters help to stabilize the map by broadening the attractor, but have not as strong impact on the results as the time delay ($\Delta$) parameter has.

Sometimes the iteration of the map generates periodic or multiperiodic signals, converges into a fixed point or becomes unstable. A slight change in the parameter can switch between these possible outcomes. 

We consider the reconstruction successful if we gather a significant number of chaotic synthetic signals that cover a well-defined area in the parameter space, and the resemblance between the input data and synthetic signals is convincing. Due to the chaotic nature of the data, no objective criterion exists for the resemblance. We cannot expect correspondence between any section of the same chaotic data either, since the trajectory never repeats itself. On the other hand, an overall similarity can be visually recognized in diverse visualization of the data.

In practice, to perform a reliable comparison between the original and synthetic signals, the following realizations have been used: the time series itself, the Fourier transform (FT), and the Broomhead-King (BK) projections \citep{bk}. The latter uses singular value decomposition to visualize the phase space trajectories in an orthogonal system. In addition, we investigate the time dependence of the Fourier parameters by calculating the analytical signals \citep{gabor} of the main pulsation and the subharmonic frequency. This provides us with a fourth, more quantitative way for the comparison.

The GFR method has been carefully and successfully tested with known chaotic and non-chaotic data. Reconstruction of chaotic data can fail if the proper parameters are not found. We prevent this by using an extended parameter space. Multiperiodic data will not yield chaotic solutions. However, very complex or stochastic components on top of the data can mimic chaos for a short while, therefore it essential to use long and accurate input data, that contains the relevant variability only. If we successfully reconstruct a data set and find it to be chaotic, it indicates that chaotic dynamics can be a plausible explanation for the observed behavior.

\begin{figure}
\includegraphics[width=0.48
\textwidth]{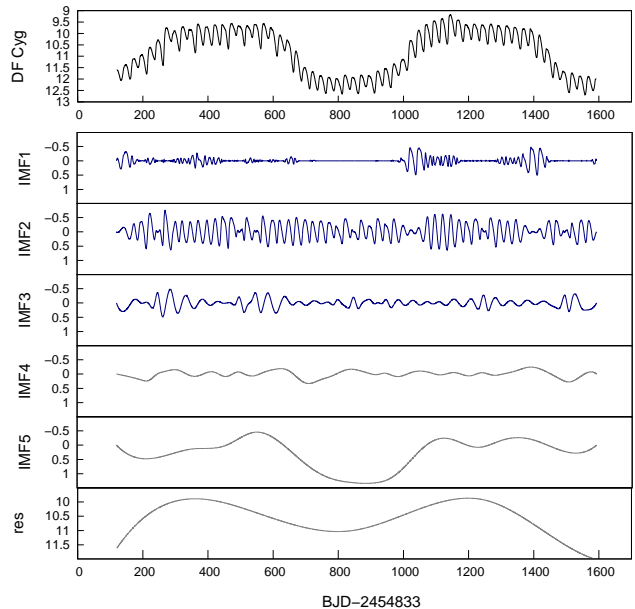}
\caption{Intrinsic mode functions of the \textit{Kepler} light curve of DF Cyg. The sum of the first three IMFs (IMF1+IMF2+IMF3) has been selected for GFR. Brightess is in \textit{Kp} magnitudes, time is in days.} 
 
\label{imfs}
\end{figure}

\subsection{The Empirical Mode Decomposition method}

The large-amplitude long-term variation represents a complication in our analysis that may prevent a robust or successful reconstruction by elevating the embedding dimension. Here we assume that the long-period RVb phenomenon is an external variation that is not connected to the pulsation dynamics, therefore we removed it before applying the GFR method. 
For this purpose we adopted the Empirical Mode Decomposition (hereafter EMD) method, that is the fundamental part of the Hilbert-Huang Transform \citep{hht}. This tool is not widely used in variable star studies, but sometimes applied in other fields of astrophysics, to study non-stationary signals \citep{Hu, Kolotov}. The EMD method was tested and found to be suitable for similar detrending application for GFR in a previous study aimed at coupled R\"ossler oscillators \citep{elim}. 

The EMD algorithm decomposes a time series into a set of intrinsic mode functions (IMFs) that show variability on different time scales with variable amplitude and frequency (see Fig. \ref{imfs}). It is based on producing spline smoothed envelopes defined by local maxima and minima and subsequent subtractions of the mean of these envelopes from the initial data. To get the first IMF, the process is repeated until the following requirements are satisfied: the number of extrema and zero-crossings should differ at most by unity and the mean of the envelopes should be zero at any point. The next IMF is obtained by subtracting the previously extracted IMF from the original data and repeating the same steps. The algorithm ends with the residual signal from which no more IMF can be extracted. We used a Python implementation of the Hilbert-Huang transform, PyHHt\footnote{https://github.com/jaidevd/pyhht}.

\section{Data preparation}
\label{dat}

\begin{figure}
\includegraphics[width=0.48
\textwidth]{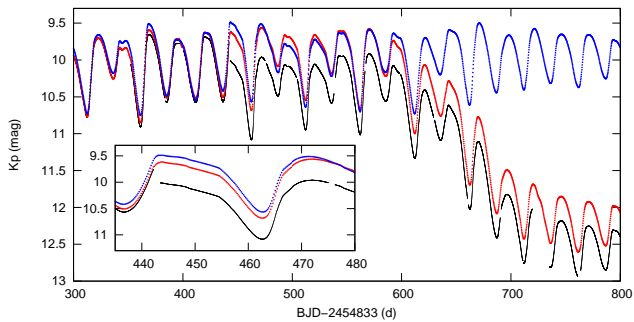}
\caption{Preparation of the light curve. A fraction of light curve shows the modifications performed before analysis. The discontinuity of the original data provided by KASOC (black) was fixed by shifting the light curves of  different observing seasons together and by filling the instrumental gaps (red). Data was resampled before long-period variation was removed (blue). The inset shows a zoom around discontinuities for better visibility.} 

\label{datamod}
\end{figure}

The large number of well-sampled cycles is a crucial criterion for the input data in our nonlinear analysis. NASA's \textit{Kepler} mission provides a high-quality, four-year-long light curve for DF Cyg containing 59 pulsation cycles and two full cycles of large-amplitude variation. As a consequence of the quarter-year rolls of the spacecraft, the light curve suffers from systematic shifts. The preparatory work for the correction of instrumental issues on the photometric solution provided by the Kepler Asteroseismic Science Operations Center (KASOC) has already been performed by \citet{bodi}. We use the same data set in our analysis with some modifications described below and showed in Fig. \ref{datamod}. 

\begin{figure*}
\includegraphics[width=1.0
\textwidth]{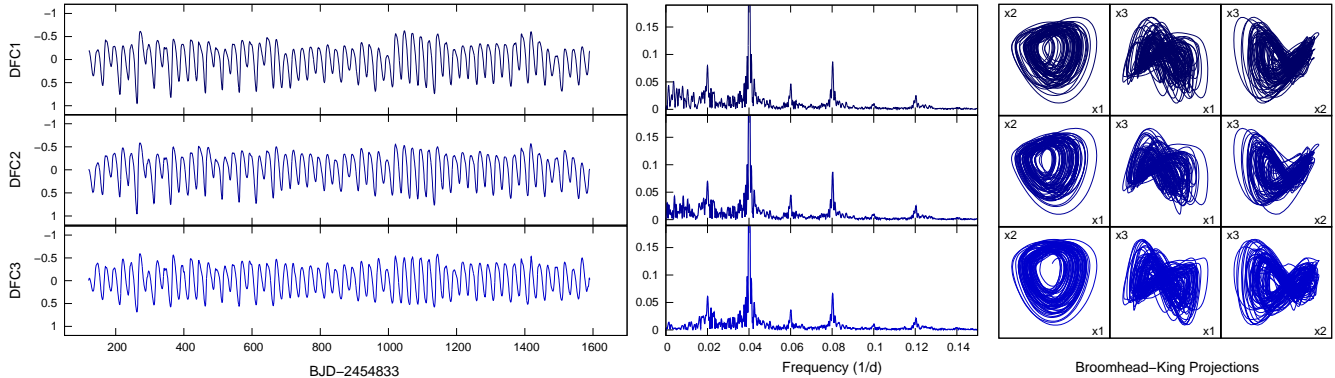}
\caption{Different realizations of the three versions of light curves (from top to bottom) that were used in GFR. Left: time series (in relative magnitude), middle: Fourier transforms, right: BK projections. The FTs only show the low-amplitude range, the main peak extends beyond the plot. } 

\label{dfcs}
\end{figure*}

GFR requires equally sampled data with an optimal number of data points, somewhere between the empirical range of 100 and 200 points per cycle. We used the K-Inpainting software, developed to fill gaps in \textit{Kepler} data \citep{inpaint1,inpaint2} to interpolate the missing points, and to resample the 30-minute cadence light curve into a data set with $\sim$5 hour sampling. Unfortunately our analysis cannot gain from the high-cadence sampling of \textit{Kepler} measurements, our tests showed that higher resolution just complicates the GFR.

The next step was the subtraction of RVb phenomenon. 
We used two different approaches to remove long-term variations: our first choice was the trigonometric polynomial fit that was used by \citet{bodi}, and as an alternative we applied the EMD method.
We produced three different versions from the resampled light curve:
\begin{enumerate}
\item for DFC1 long-term variations were removed by a trigonometric polynomial function.
\item for DFC2 we applied the EMD method. The decomposition is displayed in Fig. \ref{imfs}. The combination of the first three IMFs (i.e., IMF1+IMF2+IMF3) was selected. Here we used magnitude units.
\item for DFC3 we applied the EMD method again, but the decomposition was performed in flux units. The sum of the first three IMFs was divided by the normalized signal composed from the rest of the IMFs and the residual signal, ie., the long-term variations. The rate of the apparent amplitude attenuation is equal to the difference between the average brightness measured in the bright and faint states in flux units, according to \citet{kissbodi}. In this case we corrected the mean brightness and amplitude changes simultaneously. We then transformed our data set to magnitude units for the reconstruction. As Fig.~\ref{dfcs} shows, this curve turned out to be more smooth than DFC2 or DFC1.
\end{enumerate}

The time series, Fourier spectra and BK projections of the three versions of the light curves are shown in Fig.~\ref{dfcs}. Some differences can be recognized: traces of the long-period RVb variations appear in DFC1, while the low-frequency region of DFC3, below 0.01 d$^{-1}$, is almost completely empty. Nevertheless, since we do not know the exact shape of light variation caused by the RVb phenomenon, we cannot decide which version of the light curve captures the pulsation signal best.

\section{Analysis and results}
\label{results}

\begin{figure*}
\includegraphics[width=1.0
\textwidth]{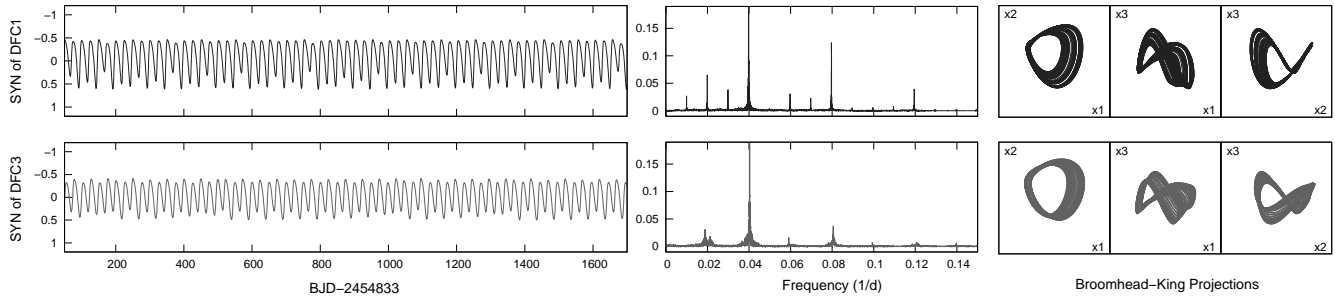}
\label{syn13}
\caption{Different realizations of the best synthetic signal examples from the reconstructions of DFC1 and DFC3. (Reconstruction parameters are: $\Delta$=10, $\xi$=0.007, $\sigma$=0, $d_e$=6,  $D_L$=2.199 and $\Delta$=26, $\xi$=0.001, $\sigma$=0.01, $d_e$=5, $D_L$=2.421.)} 
\end{figure*}

\begin{figure*}
\includegraphics[width=1.0
\textwidth]{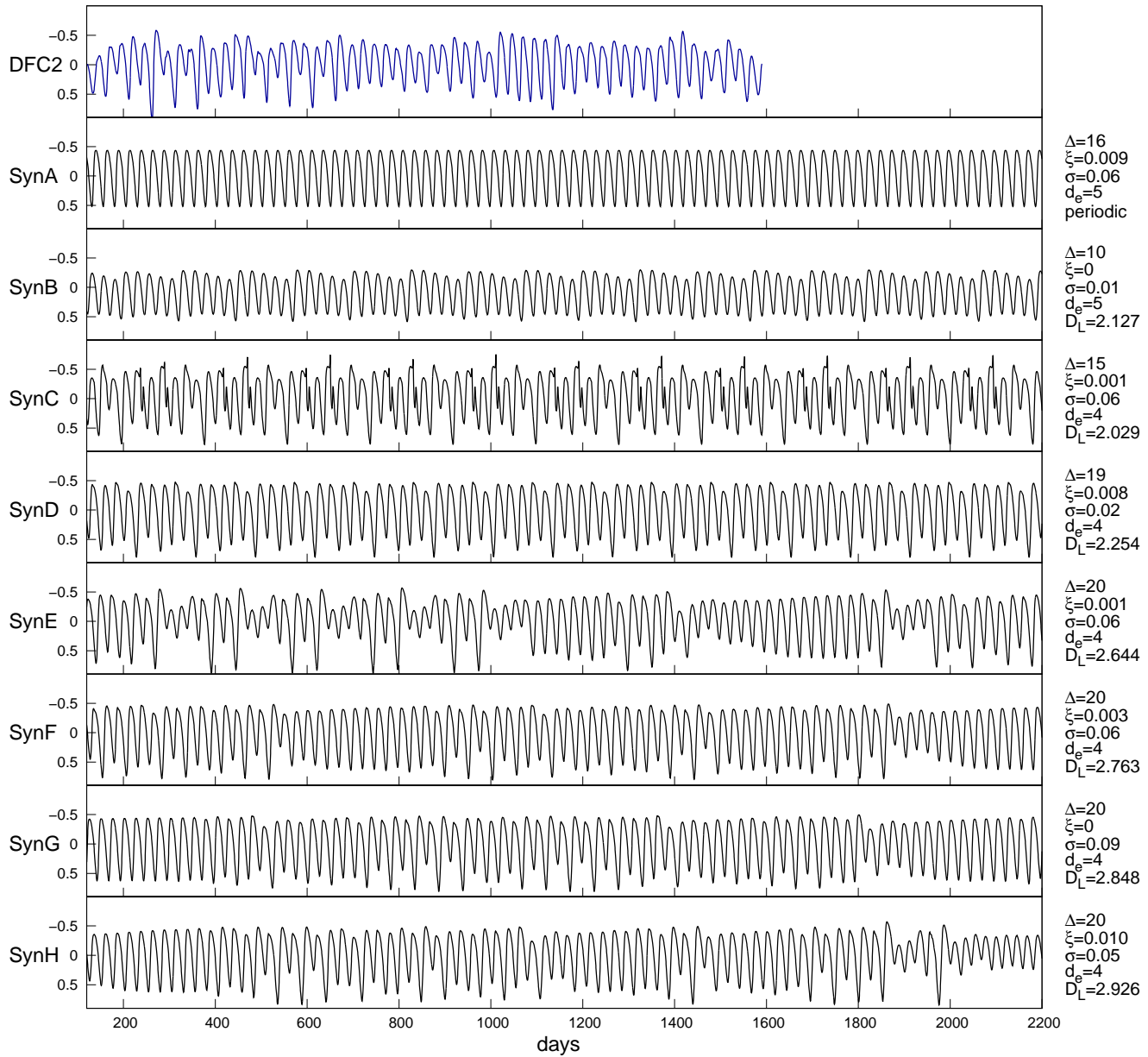}
\caption{Examples of synthetic signals and their parameters from the reconstruction of DFC2.} 
\label{results1}
\end{figure*}

\begin{figure*}
\includegraphics[width=1.0
\textwidth]{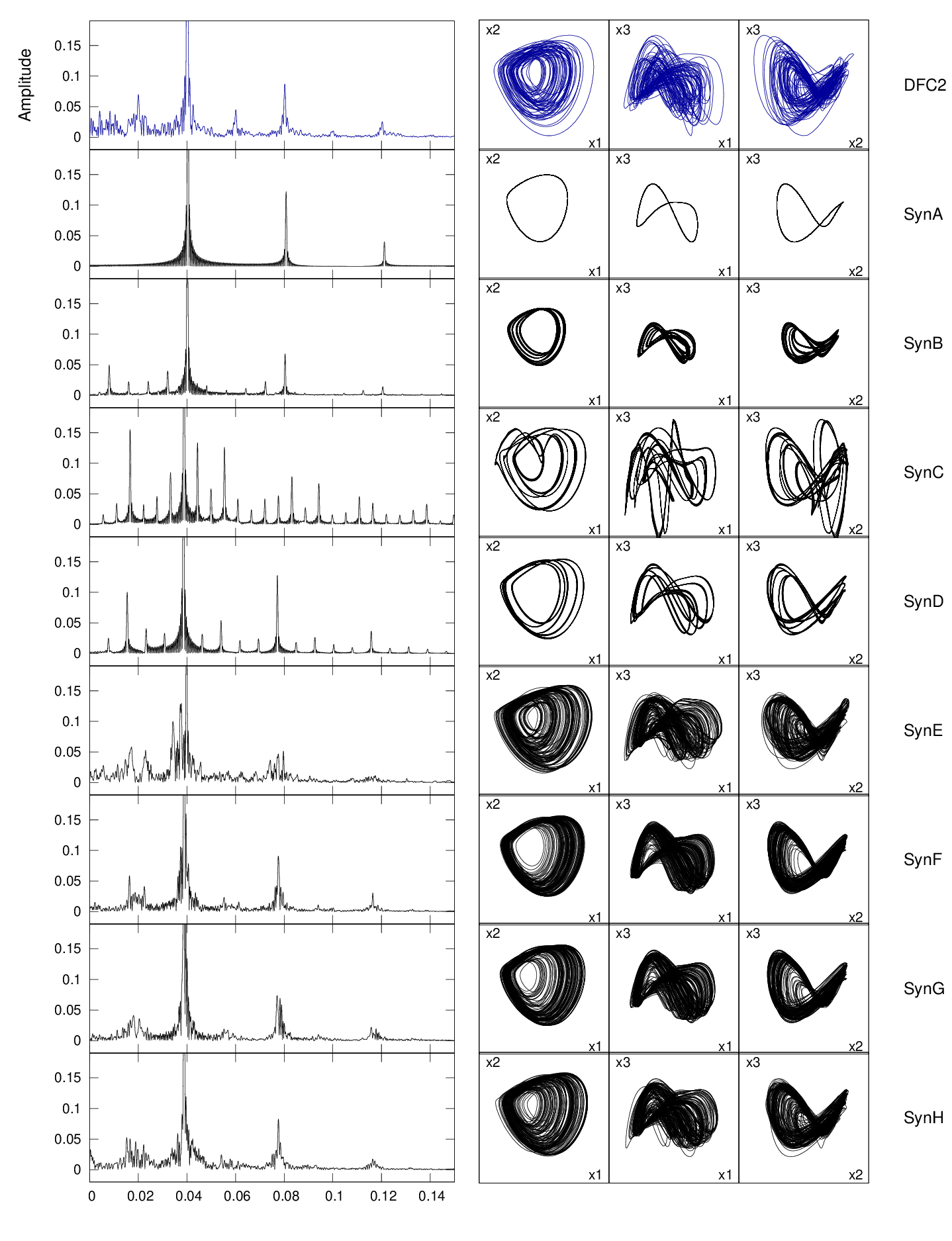}
\caption{FTs and BK projections of synthetic signals presented in Fig.~\ref{results1}.} 
\label{results2}
\end{figure*}

\begin{figure}
\includegraphics[width=0.5
\textwidth]{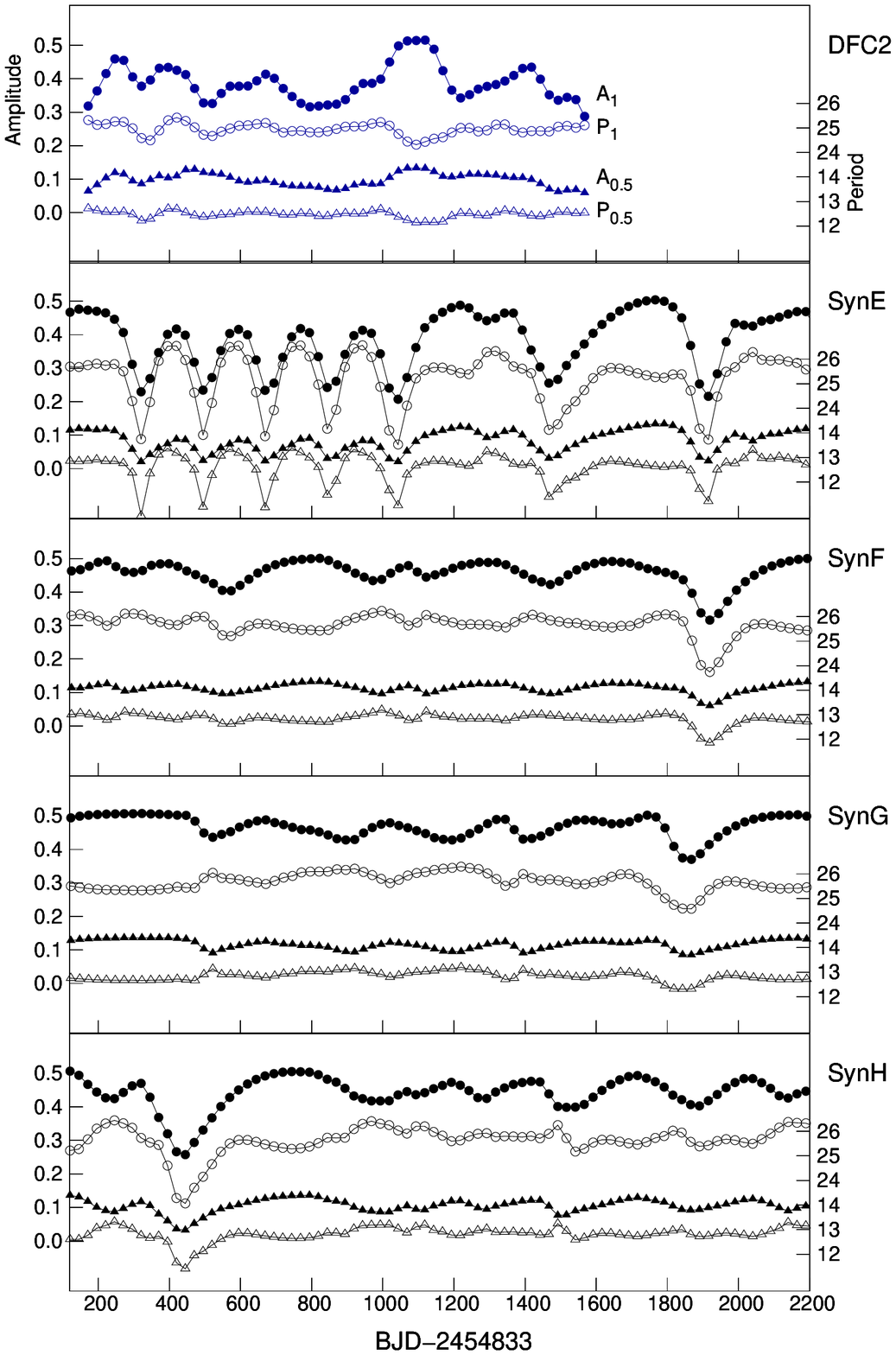}
\caption{Analytical functions of DFC2 and the best synthetic signals from its reconstruction: amplitude (filled symbols) and period (empty symbols) changes of the main oscillation(circles) and the subharmonic frequency (triangles).}  
\label{results3}
\end{figure}

We performed GFR for all three light curve versions with the same settings. The parameters were set as follows: $\Delta=4,5,6\dots30$ (time delay), $\xi=n\,0.0001$ (noise intensity) $\sigma$=$n\,0.001$ (smoothing parameter) where $n=1,2,3\dots10$. This large parameter space was used in our previous studies and was found to be extended enough to find large numbers of chaotic signals, whenever there were any. We performed GFR in all embedding dimensions implemented: $d_e$=4, 5 and 6. We iterated synthetic signals up to 300 cycles. We note here that the artificial noise and smoothing distort the data much less than the detrending techniques aimed at the elimination of RVb phenomenon, trajectories are modified within the width of lines displayed in Fig.~\ref{dfcs}. We present our findings concerning the three version of light curve below: 

\begin{enumerate}
\item We found 170 chaotic synthetic light curves as a result of the reconstruction of DFC1. None of the synthetic light curves resembled DFC1 sufficiently when the reconstruction was performed in 4- or 5-dimensional embedding space. However, a few synthetic signals in $d_e$=6 resemble the strongly alternating part of DFC1, which is typical in the first 22 cycles of the observed light curve. 
These signals have Lyapunov dimensions between 2.199--2.265. The best synthetic light curve from the reconstruction of DFC1 is presented in the upper panels of Fig.~\ref{syn13}

\item The GFR of DFC2 yielded 627 chaotic synthetic signals. 
The higher number of chaotic signals indicates that the reconstruction of DFC2 was more successful than that of DFC1. Among these signals we found numerous that show good resemblance to DFC2, especially at $\Delta$=20, in each embedding space. We present this reconstruction in more detail: in Fig.~\ref{results1} we show examples of 'bad' and 'good' synthetic signals. SynA is a periodic solution, while SynB, SynC and SynD are chaotic signals but different from DFC2. The overall difference is more prominent in the Fourier Transforms and BK projections in Fig.~\ref{results2}. The periodic signal appears in the FT as a single frequency peak with a set of harmonics, and a limit cycle in the BK projections. SynE, SynF, SynG, and SynH were all reconstructed with $\Delta$=20, we found that these signals nicely resemble DFC2, their trajectories explore almost the same extent of the phase space. However, SynE seems to have more violent cycle-to-cycle changes, which can be also seen in the analytical functions of Fig.~\ref{results3} that displays the amplitude and period change of the fundamental and the subharmonic oscillations. These plots show that the amplitudes of the fundamental and subharmonic frequency correlate, and the same is true for the period changes. On the other hand, the amplitude seems to correlate with the period for some time, then it seems to anticorrelate. This behaviour is also seen at SynF, SynG and SynH. Furthermore, the magnitude and time-scale of the amplitude and period variations are also similar that of DFC2. $D_L$ values of the best synthetic signals of these reconstruction (similar to SynF, SynG, and synH) were calculated to be $\sim$2.8 (2.720--2.926).

\item In the case of DFC3 we obtained 425 chaotic signals, but could not discover as strong resemblance among the synthetic signals as in the case of DFC2. An example synthetic signal is displayed in the lower panels of Fig.~\ref{syn13}. The typical $D_L$ of the these signals are $\sim$2.4 (2.358--2.421).   
\end{enumerate}

We conclude that the reconstruction of DFC2 was the most successful as it reproduced many characteristics, thus we can adopt $D_L$=$\sim$2.8 for the fractal dimension of the dynamics of the pulsation in DF Cyg. $D_L$ values calculated in the other two reconstructions ($\sim$2.2 and $\sim$2.4 in the cases of DFC1 and DFC3) probably underestimate the true value because they capture only some simpler features. We note that the GFR method is able to calculate Lyapunov exponents with high accuracy and $D_L$ with three digits, but the scatter in the values for the best synthetic signals prevents us from estimating $D_L$ with more than one decimal place accuracy. 

\section{Summary and conclusions} 
\label{sum}

We attempted the global flow reconstruction of an RVb-type variable star for the first time. Our method required to separate the long-term variability from the pulsation. The elimination is not unambiguous, therefore we used different techniques, and investigated three different versions of the light curve.
When RVb variations were subtracted with a trigonometrical polynomial, only the amplitude alternation could be reconstructed. We suspect that the reconstruction failed because the long-term variation was not removed properly. 
When we used the EMD method to eliminate the long-term variation from the magnitude data, we were able to successfully reconstruct the light curve, and calculated $D_L$ to be $\sim$2.8. This result suggests that the light variation of DF Cyg can be explained with the presence of low-dimensional, deterministic chaos in the pulsation. 
In the third case we applied the EMD method to the flux light curve. We assumed that scaling the stellar flux with the long-period variation would provide equivalent or better results than subtraction in magnitudes. However, the complex behavior that typifies DF Cyg was missed. It is likely that the fractal dimensions calculated in these reconstructions ($D_L \approx 2.4$) are underestimations.

Chaotic behavior can be the result of energy exchange between two oscillations. The fingerprint of this process can be found by the linear stability analysis of the map at its fixed points. For R Sct, the linear stability analysis of the fixed points revealed a 2:1 resonance between two extremely nonadiabatic modes. On the other hand, the properties of fixed points could not be determined for AC Her because phase space trajectories did not explore the vicinity of the fixed points, i.e., the amplitude of light curve never became very small. Our linear stability analysis of DF Cyg led to the same result as of AC Her: no resonance could be identified as the origin of chaotic dynamics.  In the high-luminosity, strongly dissipative hydrodynamical model sequence 
published by \citet{moskalik1990}, a half-integer resonance (the 5:2 resonance between the fundamental and second overtone mode) initiated a period doubling cascade to chaos in the 9.5-16 d period range. The same resonance caused bifurcations with different properties in the lower luminosity sequence that eventually ended in a period-one cycle at $\sim21.1$ days. These calculations also showed that the strong dissipation and nonlinearity causes considerable shifts in the resonances, therefore we cannot simply extrapolate to the longer period regime of DF Cyg based on those models.

Irregularities become more prevalent towards longer periods in RV Tau stars. However, there are exceptions, some long-period RV Tau stars seem to be less irregular than shorter-period ones. The Lyapunov dimension is used to quantify chaos, thus it is a good measure of the rate of irregularity, if it comes from chaotic dynamics. \citet{kollath1998} compared  the reconstructions of W Virginis models, AC Her, and R Sct and noticed that Lyapunov dimensions seem to increase with the pulsation period. DF Cyg does not fit into the trend, its period being shorter than that of AC Her ($\sim$37.7 d), but our analysis revealed a higher Lyapunov dimension. This could suggests that the relationship between the pulsation period and the irregularity is not unambiguous. Alternatively, the origin of the discrepancy could be that DF Cyg is an RVb variable. If small-scale variations are present in the obscuring disk, they could introduce additional variations to the light curve that we did not account for, increasing the Lyapunov dimension. Finally, AC Her itself may be more regular than other RV Tau stars. We note that the reconstruction of AC Her was less robust than that of R Sct, and the estimated Lyapunov dimension had a large uncertainty too. 

In order to understand the putative relation between periods and Lyapunov dimensions, more RV Tau stars need to be analyzed with nonlinear methods. With the increase of data quality and length of observations hopefully this could be achieved in the near future.  

Finally, we remark that while the observations of DF~Cyg (as well as of R Sct and AC Her) can be explained with chaotic dynamics, the confirmation of the existence of chaos in RV Tau stars with non-linear hydrodynamic models is yet to come.

\section*{Acknowledgments}

This project has been supported by the LP2014-17 and the  LP2018-7/2018 Programs of the Hungarian Academy of Sciences, and by the NKFIH K-115709 and PD-121203 grants of the Hungarian National Research, Development and Innovation Office. E.P.\ was supported by the J\'anos Bolyai Research Scholarship of the Hungarian Academy of Sciences. Funding for the \textit{Kepler} mission is provided by the NASA Science Mission directorate.  The Kepler Team and the Kepler Guest Observer Office
are recognized for helping to make these data possible. Fruitful discussions with Dr.~L\'aszl\'o Moln\'ar are gratefully acknowledged. We thank the referee for their comments that helped to improve this paper.

\label{lastpage}

\end{document}